\documentclass[12pt]{article}
\usepackage{amsmath,amssymb}
\numberwithin{equation}{section}

\title{{\bf Trace distance from the viewpoint of quantum operation
techniques}}

\author{Alexey E. Rastegin}
\date{\small Department of Theoretical Physics, Irkutsk State
University, Gagarin Bv. 20, Irkutsk 664003, Russia}

\begin{document}
\maketitle
\begin{abstract}
In the present paper, the trace distance is exposed within the quantum
operations formalism. The definition of the trace distance in terms
of a maximum over all quantum operations is given. It is shown that
for any pair of different states, there are an uncountably infinite
number of maximizing quantum operations. Conversely, for any
operation of the described type, there are an uncountably infinite
number of those pairs of states that the maximum is reached by the
operation. A behaviour of the trace distance under considered
operations is studied. Relations and distinctions between the trace
distance and the sine distance are discussed.

\vspace{3mm}
03.67.-a, 03.65.Ta, 02.10.Yn
\end{abstract}

%\vspace{5mm}

\protect\section{Introduction}

The formalism of quantum operations provides a unified treatment of
possible state change in quantum theory \cite{kraus1,nchuang}. The
key results on the subject of quantum operations have their origins
in papers by Hellwig and Kraus \cite{hell1,hell2}, by Kraus
\cite{kraus2}, by Lindblad \cite{lind} and by Choi \cite{choi}. The
two basic transformations, the unitary evolution and the projective
measurement, are the simplest examples of quantum operations. But very
different operations are just needed in quantum information
processing. For example, we consider distinguishing two
non-orthogonal states. This task arises in the quantum cryptography
protocol B92 \cite{bennett} and binary optical communication
\cite{paris}. The well-known scheme proposed by Helstrom
\cite{helstrom} is not error-free (except the case of orthogonality).
Nevertheless, if we allow inconclusive answers then probabilistic
error-free distinction is possible \cite{ivan,dieks,peres1}. This
scheme is usually referred to as {\it unambiguous discrimination}
\cite{chef1,chef2,turner}. The non-orthogonality of states to be
distinguished means that no projective measurement can hit. Here we
must look to generalized measurements \cite{peres1,peres2}. As it is
shown in \cite{chef1,chef2}, a rigorous treatment of arbitrary number
of those signals that should be discriminated is naturally dealt
within the quantum operations techniques.

In the light of those topics that are the subject of active research,
the techniques of quantum operations gain significance of standard
powerful tool. Indeed, many important protocols can be recast as
special cases of quantum operation; for instance, the broadcasting
\cite{barnum}, the teleportation \cite{caves}, the state
separation \cite{barn1,qiu1,feng}, and the procedure that
interpolates between unambiguous discrimination and the Helstrom
scheme \cite{jezek}. A model of computations with mixed states is
formally posed in terms of trace-preserving quantum operations
\cite{aharonov}. So, it is utmost importance that we should have an
operational meaning of basic notions of quantum theory. The revision
of needed background within the quantum operations formalism may
provide a new viewpoint on the habitual concepts. The aim of the
present work is to give a combined exposition of the trace distance
and the quantum operations in one location. We will also discuss
concerning questions.

The paper is organized as follows. In the remainder of this
section, we briefly recall necessary tools of quantum operations
techniques. In Section 2 we offer a non-standard definition of the
trace distance. Due to this new definition, a certain subclass of
quantum operations will be specified. Each of these operations
maximizes a difference between two probabilities that are
generated by the operation on some pairs of inputs. For given pair
of inputs, there is uncountably set of such maximizing operations.
On other hand, for any quantum operation of specified type, there
is uncountably set of input pairs with described property. In
Section 3 a change of the trace distance under the maximizing
operation is examined. If an operation maximizes difference
between probabilities generated by inputs then the trace distance
between outputs is bounded above. We also discuss statistical
properties of this change of the trace distance. In Section 4
relations of the trace distance to the sine distance are
considered. The bounds on the maximum of difference between these
distances are given. Section 5 concludes the paper with a summary
of obtained results.

Let ${\cal{H}}_1$ and ${\cal{H}}_2$ be the finite-dimensional Hilbert
spaces. In general, these spaces are assumed to be different. To mark
distinction of spaces, we shall supply the item of trace operation by
a label. That is, the trace ${\rm{tr}}_1\{\cdot\}$ is taken over
${\cal{H}}_1$, the trace ${\rm{tr}}_2\{\cdot\}$ is taken over
${\cal{H}}_2$. Consider any process ${\cal{E}}$ that leads to a map
\begin{equation}
\rho\to\rho':=\frac{{\cal{E}}(\rho)}{{\rm{tr}}_2\{{\cal{E}}(\rho)\}}
\ , \label{eq1}
\end{equation}
where an input $\rho$ is some normalized state on ${\cal{H}}_1$ and
an output $\rho'$ is some normalized state on ${\cal{H}}_2$. If this
map is consistent with the laws of quantum theory, then ${\cal{E}}$
is a {\it quantum operation} with the input space ${\cal{H}}_1$ and
the output space ${\cal{H}}_2$ \cite{nchuang}. The normalizing
divisor in (\ref{eq1}) is the probability that the above process
occurs. So we demand that
\begin{equation}
0\leq{\rm{tr}}_2\{{\cal{E}}(\rho)\}\leq1 \label{eq2}
\end{equation}
for each input $\rho$. In addition, a map ${\cal{E}}$ must be linear
and completely positive \cite{nchuang}.

The operator-sum representation is a key result of the quantum
operations formalism. Namely \cite{kraus1,nchuang}, the map
${\cal{E}}$ is a quantum operation if and only if
\begin{equation}
{\cal{E}}(\rho)=\sum\nolimits_{\mu} {\mathsf{E}}_{\mu}
\,\rho\, {\mathsf{E}}_{\mu}^{\dagger} \label{eq3}
\end{equation}
for some set of operators $\{{\mathsf{E}}_{\mu}\}$. These operators
map the input space ${\cal{H}}_1$ to the output space ${\cal{H}}_2$.
Some features of given quantum operation are determined by properties
of the positive operator
\begin{equation}
{\mathbf{T}}:=
\sum\nolimits_{\mu} {\mathsf{E}}_{\mu}^{\dagger}
{\mathsf{E}}_{\mu}
\ . \label{eq4}
\end{equation}
In the following, we will essentially use the equality
\begin{equation}
{\rm{tr}}_2\bigl\{{\cal{E}}(\rho)\bigr\}=
{\rm{tr}}_1\bigl\{{\mathbf{T}}\rho\bigr\}
 \ . \label{eq5}
\end{equation}
This is based on the operator-sum representation and the properties
of the trace. Suppose ${\mathsf{A}}:{\cal{H}}_1\to{\cal{H}}_2$ and
${\mathsf{B}}:{\cal{H}}_2\to{\cal{H}}_1$ are linear operators. Then
by the cyclic property we have
${\rm{tr}}_2\{{\mathsf{AB}}\}={\rm{tr}}_1\{{\mathsf{BA}}\}$. Tracing
the right-hand side of (\ref{eq3}) and using the cyclic property
and the linearity of the trace, we at once obtain (\ref{eq5}).
The inequality (\ref{eq2}) must be satisfied for all inputs.
Combining this with (\ref{eq5}), we get
${\mathbf{0}}\leq{\mathbf{T}}\leq{\mathbf{1}}$.

We shall also use the fact \cite{nchuang} that operator
$(\rho-\varrho)$ can be represented as
$\rho-\varrho={\mathbf{Q}}-{\mathbf{R}}$, where ${\mathbf{Q}}$ and
${\mathbf{R}}$ are positive operators with the orthogonal support
spaces.  [Recall that {\it support} of an operator is defined as the
vector space orthogonal to its kernel.] Indeed, due to the spectral
decomposition of $(\rho-\varrho)$ we obtain
\begin{align}
{\mathbf{Q}}&:=\sum\nolimits_{q} \lambda_q
\> |q\rangle\langle q| \ ,
\label{eq6} \\
{\mathbf{R}}&:=\sum\nolimits_{r} \varkappa_r
\> |r\rangle\langle r| \ ,
\label{eq7}
\end{align}
where the $\lambda_q$'s and the $(-\varkappa_r)$'s are strictly
positive and strictly negative eigenvalues of operator
$(\rho-\varrho)$ respectively. Let ${\rm supp}({\mathsf{A}})$ denote
the support of an operator ${\mathsf{A}}$. Then the input space
${\cal{H}}_1$ can be expressed as
\begin{equation}
{\cal{H}}_1={\rm supp}({\mathbf{Q}})\oplus {\rm
supp}({\mathbf{R}})\oplus{\cal{K}} \ ,\label{eq8}
\end{equation}
where ${\rm supp}({\mathbf{Q}})$ is spanned by $|q\rangle$'s,
${\rm supp}({\mathbf{R}})$ is spanned by $|r\rangle$'s and
${\cal{K}}$ denotes the kernel of operator $(\rho-\varrho)$.

\protect\section{Non-standard definition}

In this section, we shall introduce a non-standard definition of trace
distance and investigate those questions that are risen in the planned
way. With each quantum operation ${\cal{E}}$, one can associate some
distance measure for quantum states. Let $\rho$ and $\varrho$ be the
normalized states on ${\cal{H}}_1$. Two positive numbers
${\rm{tr}}_2\{{\cal{E}}(\rho)\}$ and
${\rm{tr}}_2\{{\cal{E}}(\varrho)\}$ give the probabilities that the
represented process occurs when $\rho$ and $\varrho$ were
respectively taken as initial state. It is natural to measure a
closeness of these states by the difference between the corresponding
probabilities.

{\bf Definition 1.} {\it Let ${\cal{E}}$ be a quantum operation. The
${\cal{E}}$-distance $d_{\cal{E}}(\rho,\varrho)$ between normalized
states $\rho$ and $\varrho$ is defined by}
\begin{equation}
d_{\cal{E}}(\rho,\varrho):=
\left|\, {\rm{tr}}_2\bigl\{{\cal{E}}(\rho)\bigr\}
- {\rm{tr}}_2\bigl\{{\cal{E}}(\varrho)\bigr\}
\right| \ . \label{eq9}
\end{equation}

It is clear that $0\leq d_{\cal{E}}\leq1$, that if
$\rho=\varrho$ then $d_{\cal{E}}(\rho,\varrho)=0$, and that
$d_{\cal{E}}$ is a symmetric function of inputs. The absolute value
of sum does not exceed the sum of absolute values so that
$d_{\cal{E}}(\rho,\varrho)\leq
d_{\cal{E}}(\rho,\omega)+d_{\cal{E}}(\omega,\varrho)$,
i.e. the triangle inequality holds. So ${\cal{E}}$-distance
obeys all the properties of a metric except only one. Namely, even if
$\rho\not=\varrho$ the equality $d_{\cal{E}}(\rho,\varrho)=0$ can
still be valid (when ${\rm dim}({\cal{H}}_1)>2$). Indeed, due
to (\ref{eq5}) the last equality is equivalent to
${\rm{tr}}_1\bigl\{{\mathbf{T}}(\rho-\varrho)\bigr\}=0$ that is
provided by ${\rm supp}({\mathbf{T}})\subseteq{\cal{K}}$. [Only in
two-dimensional input space, ${\cal{E}}$-distance is a
metric because $\rho\not=\varrho$ implies here that
${\rm dim}({\cal{K}})=0$ is inevitable.] It is unfit that
$d_{\cal{E}}(\rho,\varrho)=0$ does not imply $\rho=\varrho$. But this
lack is repaired by the maximization over all quantum operations. It
turns out that such an approach leads to well-known metric on quantum
states, namely to the trace distance.

Let $|{\mathsf{A}}|$ denote the positive square root of
${\mathsf{A}}^{\dagger}{\mathsf{A}}$ (for any positive operator
there exists a unique positive square root \cite{reed}). The trace
distance between states $\rho$ and $\varrho$ is traditionally
defined by \cite{nchuang}
\begin{equation}
D(\rho,\varrho):=\frac{1}{2}\ {\rm{tr}}_1 |\rho-\varrho| \ .
\label{eq10}
\end{equation}
The trace distance is simply expressed in terms of operators
${\mathbf{Q}}$ and ${\mathbf{R}}$ \cite{nchuang}. Since the
supports of these operators are orthogonal, we have
$|{\mathbf{Q}}-{\mathbf{R}}|={\mathbf{Q}}+{\mathbf{R}}$ and
\begin{equation}
D(\rho,\varrho)=\frac{1}{2}\ {\rm{tr}}_1({\mathbf{Q}})
+\frac{1}{2}\ {\rm{tr}}_1({\mathbf{R}})\ . \label{eq11}
\end{equation}
When states $\rho$ and $\varrho$ are normalized to the unit trace, the
right-hand side of (\ref{eq11}) is equal to
${\rm{tr}}_1({\mathbf{Q}})={\rm{tr}}_1({\mathbf{R}})$. The trace
distance has many attractive properties that makes it a proper
measure of closeness of quantum states (for a discussion, see
subsection 9.2.1 of reference \cite{nchuang}). The mentioned
connection between the ${\cal{E}}$-distance and the trace distance is
established by the following statement.

{\bf Theorem 1.} {\it For any normalized states $\rho$ and $\varrho$,}
\begin{equation}
\underset{\cal{E}}{\max} \ d_{\cal{E}}(\rho,\varrho)
=D(\rho,\varrho) \ , \label{eq12}
\end{equation}
{\it where maximum is taken over all quantum operations ${\cal{E}}$.
The maximum is reached by quantum operation ${\cal{E}}$ if and only
if operator ${\mathbf{T}}$ is equal to either the projector onto
${\rm supp}({\mathbf{Q}})$ or the projector onto
${\rm supp}({\mathbf{R}})$, up to additive term ${\mathbf{M}}$
satisfying ${\rm supp}({\mathbf{M}})\subseteq{\cal{K}}$ and
${\mathbf{0}}\leq{\mathbf{M}}\leq{\mathbf{1}}$.}

{\bf Proof.} We shall now suppose that $\rho\not=\varrho$ (otherwise
both distances are zero, ${\cal{K}}={\cal{H}}_1$ and the statement of
theorem does not add anything new). Then both sets $\{\lambda_q\}$ and
$\{\varkappa_r\}$ are nonempty. Due to (\ref{eq5}) we have
\begin{equation}
d_{\cal{E}}(\rho,\varrho)=
\bigl|{\rm{tr}}_1\{{\mathbf{TQ}}\}
- {\rm{tr}}_1\{{\mathbf{TR}}\}
\bigr| \ . \label{eq13}
\end{equation}
Since operators ${\mathbf{Q}}$ and ${\mathbf{R}}$ are positive and
${\mathbf{0}}\leq{\mathbf{T}}\leq{\mathbf{1}}$, each of two traces in
the right-hand side of (\ref{eq13}) is nonnegative and no greater
than
$D(\rho,\varrho)={\rm{tr}}_1({\mathbf{Q}})={\rm{tr}}_1({\mathbf{R}})$.
So ${\cal{E}}$-distance between states $\rho$ and $\varrho$
does not exceed the trace distance between them. The equality is
reached in two cases: (i)
${\rm{tr}}_1\{{\mathbf{TQ}}\}={\rm{tr}}_1({\mathbf{Q}})$
and ${\rm{tr}}_1\{{\mathbf{TR}}\}=0$; (ii)
${\rm{tr}}_1\{{\mathbf{TQ}}\}=0$ and
${\rm{tr}}_1\{{\mathbf{TR}}\}={\rm{tr}}_1({\mathbf{R}})$.
We shall consider the case (i) only; the case (ii) follows the same
pattern. If ${\mathbf{T}}$ is the sum of projector onto
${\rm supp}({\mathbf{Q}})$ and some ${\mathbf{M}}$ with
${\rm supp}({\mathbf{M}})\subseteq{\cal{K}}$ then the conditions of
the case (i) take place. Suppose now that the conditions of the case
(i) are fulfilled. Let the $|a\rangle$'s form an orthonormal set in
${\rm supp}({\mathbf{Q}})\oplus{\cal{K}}$. Clearly, $\langle
a|r\rangle=0$ for all $a$ and $r$. Then operator ${\mathbf{T}}$ can
be expressed by
\begin{equation}
{\mathbf{T}}=\sum\nolimits_a c_{aa} |a\rangle\langle a|+
\sum\nolimits_{ar}\bigl(c_{ar}|a\rangle\langle r|
+c_{ra}|r\rangle\langle a|\bigr)
+\sum\nolimits_r c_{rr} |r\rangle\langle r|
\ , \label{eq14}
\end{equation}
where all the diagonal elements lie in the interval $[0;1]$. Because
the $\varkappa_r$'s in (\ref{eq6}) are strictly positive, the
condition ${\rm{tr}}_1\{{\mathbf{T}}{\mathbf{R}}\}=0$ implies that
$c_{rr}=0$ for all values of label $r$ (so the kernel of
${\mathbf{T}}$ is not zero-dimensional). Moreover, all the
off-diagonal elements $c_{ar}$ and $c_{ra}$ are also zero. To prove
this fact, we use a modification of the method of reference
\cite{rast1}. Let us fix the values of $a$ and $r$, and let us
consider a subspace ${\rm span}\{|a\rangle,|r\rangle\}$. In this
subspace, the action of ${\mathbf{T}}$ is described by the matrix
\begin{equation}
\left(\begin{array}{cc}
 c_{aa} & \alpha-i\beta \\
 \alpha+i\beta & 0
\end{array}\right)
\ . \label{eq15}
\end{equation}
Here $\alpha$ and $\beta$ are real, and
$c^*_{ar}=c_{ra}=\alpha+i\beta$. Due to positivity of ${\mathbf{T}}$,
both eigenvalues of the matrix (\ref{eq15}) are nonnegative. This is
valid if and only if $\alpha=\beta=0$ and therefore
$c_{ar}=c_{ra}=0$. Thus, only the first sum in the right-hand side of
(\ref{eq14}) is nonzero, whence
${\rm{supp}}({\mathbf{T}})\subseteq{\rm{supp}}({\mathbf{Q}})\oplus{\cal{K}}$.
Let the $|b\rangle$'s form an orthonormal basis in ${\cal{K}}$.
Obviously, $\langle q|b\rangle=0$ for all $q$ and $b$. Then operator
${\mathbf{T}}$ can be represented as
\begin{equation}
{\mathbf{T}}=\sum\nolimits_q t_{qq} |q\rangle\langle q|+
\sum\nolimits_{qb}\bigl(t_{qb}|q\rangle\langle b|
+t_{bq}|b\rangle\langle q|\bigr)
+\sum\nolimits_b t_{bb} |b\rangle\langle b|
\ . \label{eq16}
\end{equation}
As before, all the diagonal elements lie in the interval $[0;1]$.
Since the $\lambda_q$'s in (\ref{eq6}) are strictly positive, the
condition
${\rm{tr}}_1\{{\mathbf{T}}{\mathbf{Q}}\}={\rm{tr}}_1({\mathbf{Q}})$
implies that $t_{qq}=1$ for all values of label $q$. So the first sum
in the right-hand side of (\ref{eq16}) must be the projector onto
${\rm{supp}}({\mathbf{Q}})$. Fixing some values of $q$ and $b$, we
shall now consider the action of ${\mathbf{T}}$ in the two-dimensional
subspace ${\rm span}\{|q\rangle,|b\rangle\}$. This action is
described by the matrix
\begin{equation}
\left(\begin{array}{cc}
 1 & \gamma-i\delta \\
 \gamma+i\delta & t_{bb}
\end{array}\right)
\ . \label{eq17}
\end{equation}
Here $\gamma$ and $\delta$ are real, and
$t^*_{qb}=t_{bq}=\gamma+i\delta$. By ${\mathbf{T}}\leq{\mathbf{1}}$
both eigenvalues of the matrix (\ref{eq17}) are no greater than 1.
This is valid if and only if $\gamma=\delta=0$ and therefore
$t_{qb}=t_{bq}=0$. Let us denote the third sum in the right-hand side
of (\ref{eq16}) by ${\mathbf{M}}$. It is obvious that this operator
satisfies ${\rm{supp}}({\mathbf{M}})\subseteq{\cal{K}}$ and
${\mathbf{0}}\leq{\mathbf{M}}\leq{\mathbf{1}}$. Then the operator
${\mathbf{T}}$ is the sum of projector onto
${\rm{supp}}({\mathbf{Q}})$ and ${\mathbf{M}}$. $\blacksquare$

The left-hand side of (\ref{eq12}) can fruitfully be considered
as a non-standard definition of the trace distance. The usual
definition was seemingly inspired on the analogy of classicality (for
details, see subsection 9.2.1 of reference \cite{nchuang}). In
contrast, the series of arguments that leads to Theorem 1 is a
self-contained nonclassical way to approach the genuine metric on
quantum states. This way provides a kind of physical interpretation
of equation (\ref{eq10}) which is rather handy for evaluating the
trace distance. Thus, we have arrived at the following definition.

{\bf Definition 2.} ({\bf Non-standard definition of trace distance)}
{\it The trace distance $D(\rho,\varrho)$ between quantum states
$\rho$ and $\varrho$ is defined by}
\begin{equation}
D(\rho,\varrho):=\underset{\cal{E}}{\max}
\,\left|\, {\rm{tr}}_2\bigl\{{\cal{E}}(\rho)\bigr\}
- {\rm{tr}}_2\bigl\{{\cal{E}}(\varrho)\bigr\}
\right| \ . \label{def}
\end{equation}

The consistency of the new definition with the customary one is
stated by Theorem 1. In connection with the definition given by
(\ref{def}) some unexpected questions are naturally risen. New
insights into relationship of quantum operations and quantum states
will be achieved by the study of these questions. Whenever the
equality $d_{\cal{E}}(\rho,\varrho)=D(\rho,\varrho)$ is done by
quantum operation ${\cal{E}}$, we will say: "the operation maximizes
probability difference between $\rho$ and $\varrho$". We ask: How
many such quantum operations?

{\it To each pair $\{\rho,\varrho\}$ of different states assign a
family of classes labelled by integer $N>1$. The class specified
by the given value $N$ contains an uncountably infinite number of
those quantum operations that have $N$-dimensional output space
and satisfy $d_{\cal{E}}(\rho,\varrho)=D(\rho,\varrho)$.}

The claimed statement is justified as follows. Let us demand that
operator ${\mathbf{T}}$ be equal to the projector onto
${\rm{supp}}({\mathbf{Q}})$. We choose a relevant number of
vectors $|q'\rangle\in{\cal{H}}_2$ and take
${\mathsf{E}}_q=|q'\rangle\langle q|$. The only thing we must assume
about these vectors is that they are all normalized. In two and more
dimensions, there are uncountably infinite number of ways to choose
$|q'\rangle$'s. Thus, for any given value $N>1$ we can build an
uncountably infinite number of those quantum operations that maximize
probability difference between $\rho$ and $\varrho$, as claimed. The
case, in which operator ${\mathbf{T}}$ should be equal to the
projector onto ${\rm{supp}}({\mathbf{R}})$, follows the same
pattern. If ${\rm{dim}}({\cal{K}})>0$ then by choice of
${\mathbf{M}}$ we obtain an additional freedom.

We have examined the question about those quantum operations that
maximize probability difference between any prescribed two states.
It is natural to inspect things in reverse order. As Theorem 1 shows,
the specific property of considered quantum operations is that both
the unity and zero are eigenvalues of ${\mathbf{T}}$. First, the
operator ${\mathbf{T}}$ can be split into sum of projector and another
operator with orthogonal supports. Second, the kernel of
${\mathbf{T}}$ is not zero-dimensional (except $\rho=\varrho$). So we
pick out the special subclass of quantum operations. Let us begin
with given quantum operation of described type. It is easy to build
those two states that probability difference between them is
maximized by the operation. In how many ways can we make such
building?

{\it A family of classes, labelled by real
${\mathfrak{D}}\in(0;1)$, is assigned to each quantum operation
${\cal{E}}$ such that operator ${\mathbf{T}}$ has unit and zero
eigenvalues. The class specified by the given value ${\mathfrak{D}}$
contains an uncountably infinite number of those pairs
$\{\rho,\varrho\}$ that obey
$D(\rho,\varrho)=d_{\cal{E}}(\rho,\varrho)={\mathfrak{D}}$}.

The justification is simple. We choose a nontrivial subspace of the
eigen\-space corresponding to unit eigenvalue of ${\mathbf{T}}$; this
subspace is designed as ${\rm{supp}}({\mathbf{Q}})$. Then we take a
nontrivial subspace of the kernel of ${\mathbf{T}}$; that subspace is
designed as ${\rm{supp}}({\mathbf{R}})$. So, the conditions
${\rm{tr}}_1\{{\mathbf{T}}{\mathbf{Q}}\}={\rm{tr}}_1({\mathbf{Q}})$
and ${\rm{tr}}_1\{{\mathbf{T}}{\mathbf{R}}\}=0$ are provided. The
orthogonal complement of
${\rm{supp}}({\mathbf{Q}})\oplus{\rm{supp}}({\mathbf{R}})$ is clearly
designed as ${\cal{K}}$. Let the $|q\rangle$'s and the $|r\rangle$'s
be those eigenvectors of ${\mathbf{T}}$ that form orthonormal sets in
${\rm{supp}}({\mathbf{Q}})$ and ${\rm{supp}}({\mathbf{R}})$
respectively. We then take positive numbers $\lambda_q$ and
$\varkappa_r$ and define operators ${\mathbf{Q}}$ and ${\mathbf{R}}$
by (\ref{eq6}) and (\ref{eq7}), respectively. Both traces
${\rm{tr}}_1({\mathbf{Q}})$ and ${\rm{tr}}_1({\mathbf{R}})$ should be
equal to ${\mathfrak{D}}$. That is, both the $\lambda_q$'s and the
$\varkappa_r$'s sum to ${\mathfrak{D}}$. Then the trace distance
between desired quantum states will be equal to ${\mathfrak{D}}$. We
now aim to build normalized states $\rho$ and $\varrho$ satisfying
$\rho-\varrho={\mathbf{Q}}-{\mathbf{R}}$. We consider the case in
which both $\rho$ and $\varrho$ are supported on
${\rm{supp}}({\mathbf{Q}})\oplus{\rm{supp}}({\mathbf{R}})$
and diagonal with respect to the orthonormal set formed by
$|q\rangle$'s and $|r\rangle$'s. Let us define these states as
\begin{align}
\rho&:=\sum\nolimits_{q} (\lambda_q+\delta\lambda_q)
\> |q\rangle\langle q| + \sum\nolimits_{r} \delta\varkappa_r
\> |r\rangle\langle r| \ ,
\label{eq18} \\
\varrho&:=\sum\nolimits_{r} (\varkappa_r+\delta\varkappa_r)
\> |r\rangle\langle r| + \sum\nolimits_{q} \delta\lambda_q
\> |q\rangle\langle q|\ ,
\label{eq19}
\end{align}
where positive variations $\delta\lambda_q$ and $\delta\varkappa_r$
must obey
\begin{equation}
\sum\nolimits_{q} \delta\lambda_q
+\sum\nolimits_{r} \delta\varkappa_r
=1-{\mathfrak{D}} \ .
\label{eq20}
\end{equation}
So the normalization of $\rho$ and $\varrho$ is provided. Because
both sets $\{\delta\lambda_q\}$ and $\{\delta\varkappa_r\}$ are
nonempty, we have an uncountably infinite number of ways to satisfy
(\ref{eq20}), as claimed above.

We have examined a maximum of $d_{\cal{E}}(\rho,\varrho)$ for the
prescribed two states $\rho$ and $\varrho$. We shall now perform the
maximization of ${\cal{E}}$-distance over all possible states.
Consider a fixed quantum operation ${\cal{E}}$ of arbitrary type. It
turns out that the desired maximum is equal to the difference between
the maximal and minimal eigenvalues of operator ${\mathbf{T}}$. By
$\Theta$ and $\theta$ we respectively denote these maximal and
minimal eigenvalues. Then the following statement holds.

{\bf Theorem 2.} {\it For arbitrary quantum operation ${\cal{E}}$,}
\begin{equation}
\underset{\rho,\varrho}{\max}
\>d_{\cal{E}}(\rho,\varrho)=\Theta-\theta
\ , \label{eqt1}
\end{equation}
{\it where the maximum is taken over all states $\rho$ and $\varrho$.}

{\bf Proof.} A value of $d_{\cal{E}}(\rho,\varrho)$ for particular
two states $\rho$ and $\varrho$ is given by (\ref{eq13}). In this
equation two operators ${\mathbf{Q}}$ and ${\mathbf{R}}$ are uniquely
determined by the two states. So both the trace of ${\mathbf{Q}}$ and
the trace of ${\mathbf{R}}$ are equal to $D(\rho,\varrho)$. Under
these conditions we can apply the result of Lemma 1 of Appendix A. By
(\ref{eqa01}) the trace ${\rm{tr}}_1\{{\mathbf{TQ}}\}$ is no greater
than $\Theta D(\rho,\varrho)$, by (\ref{eqa02}) the trace
${\rm{tr}}_1\{{\mathbf{TR}}\}$ is no less than
$\theta D(\rho,\varrho)$. Therefore,
\begin{equation*}
d_{\cal{E}}(\rho,\varrho)\leq(\Theta-\theta)D(\rho,\varrho)
\leq\Theta-\theta \ ,
\end{equation*}
where we used $D(\rho,\varrho)\leq1$. The right-hand side of
(\ref{eqt1}) is reached under the following two conditions. The
density operator $\rho$ must be multiplied by the normalizing factor
projector onto nontrivial subspace of the eigenspace of
${\mathbf{T}}$ corresponding to eigenvalue $\Theta$; the density
operator $\varrho$ must be multiplied by the normalizing factor
projector onto nontrivial subspace of the eigenspace of
${\mathbf{T}}$ corresponding to eigenvalue $\theta$. $\blacksquare$

It is obvious that for trace-preserving quantum operation the
${\cal{E}}$-distance is equal to zero. In line with this fact, the
right-hand side of (\ref{eqt1}) vanishes because
${\mathbf{T}}={\mathbf{1}}$ for all trace-preserving operations. If
quantum operation is maximizing, then both the unity and zero are
eigenvalues of ${\mathbf{T}}$ and the right-hand side of
(\ref{eqt1}) is equal to 1. The latter is the maximal acceptable value
of ${\cal{E}}$-distance. This is another reason for usage of the word
'maximizing'.

\protect\section{Behaviour under the maximizing \\ quantum operation}

In mutual relations of quantum operations and trace distance the
following result of great moment is well known \cite{nchuang,ruskai}.
Namely, no deterministic process increases the distance between two
quantum states. That is, if ${\cal{E}}$ is a trace-preserving quantum
operation then
\begin{equation}
D({\cal{E}}(\rho),{\cal{E}}(\varrho))\leq
D(\rho,\varrho)
\label{eq21}
\end{equation}
for arbitrary normalized states $\rho$ and $\varrho$. This result is
usually referred to as {\it contractivity} of the trace distance
under trace-preserving quantum operations. According to (\ref{eq5}),
for all trace-preserving operations ${\mathbf{T}}={\mathbf{1}}$ and
therefore states ${\cal{E}}(\rho)$ and ${\cal{E}}(\varrho)$ are
normalized. The quantum operations that are the subject of interest
in the present work do not preserve the trace. Nevertheless, the
considered operations may be almost contractive in a specific sense.
As has been shown above, with each quantum operation of the
described type one can associate an uncountably infinite set of pairs
with specified property. It is for these states that the following
property of the operation is valid.

{\bf Theorem 3.} {\it If the quantum operation ${\cal{E}}$ maximizes
probability difference between normalized states $\rho$ and $\varrho$
then}
\begin{equation}
D(\rho',\varrho')\leq
p_{\rm{m}}^{-1}D(\rho,\varrho) \ ,
\label{eq22}
\end{equation}
{\it where states $\rho'$ and $\varrho'$ are normalized outputs of the
operation and $p_{\rm{m}}$ is maximum among two probabilities
${\rm{tr}}_2\{{\cal{E}}(\rho)\}$ and
${\rm{tr}}_2\{{\cal{E}}(\varrho)\}$.}

{\bf Proof.} We shall mean that $\rho\not=\varrho$ and therefore two
probabilities are different. With no loss of generality,
${\rm{tr}}_2\{{\cal{E}}(\rho)\}>{\rm{tr}}_2\{{\cal{E}}(\varrho)\}$.
This implies that the case (i) is realized (see the proof of Theorem
1). Due to (\ref{eq5}) the conditions of the case (i) can be
represented as
${\rm{tr}}_2\{{\cal{E}}({\mathbf{Q}})\}={\rm{tr}}_1({\mathbf{Q}})$
and ${\rm{tr}}_2\{{\cal{E}}({\mathbf{R}})\}=0$, whence
\begin{align}
D(\rho,\varrho) &= {\rm{tr}}_2\{{\cal{E}}({\mathbf{Q}})\}
-{\rm{tr}}_2\{{\cal{E}}({\mathbf{R}})\} \nonumber\\
&\geq {\rm{tr}}_2\{\Pi{\cal{E}}({\mathbf{Q}})\}
-{\rm{tr}}_2\{\Pi{\cal{E}}({\mathbf{R}})\} \nonumber\\
&={\rm{tr}}_2\{\Pi({\cal{E}}(\rho)- {\cal{E}}(\varrho))\}
\label{eq23}
\end{align}
for arbitrary projector $\Pi$. In the last line of (\ref{eq23})
the linearity of the trace and the map (\ref{eq3}) is used. According
to (\ref{eq1}) we further have
\begin{align}
\rho'&=p^{-1}_{\rm{m}}\ {\cal{E}}(\rho)
\ , \label{eq24} \\
\varrho'&=p^{-1}_{\rm{n}}\ {\cal{E}}(\varrho)
\ , \label{eq25}
\end{align}
where $p_{\rm{m}}={\rm{tr}}_2\{{\cal{E}}(\rho)\}$ and
$p_{\rm{n}}={\rm{tr}}_2\{{\cal{E}}(\varrho)\}$. As it is well known
(see equation (9.22) of reference \cite{nchuang}), there exists a
projector $\Pi$ such that
\begin{equation}
{\rm{tr}}_2\{\Pi(\rho'-\varrho')\}=
D(\rho',\varrho')
\ . \label{eq26}
\end{equation}
Using Eqs. (\ref{eq24}) and (\ref{eq25}) and inequality
$p_{\rm{m}}>p_{\rm{n}}$ later, the last line of  (\ref{eq23}) can
be put in the form
\begin{equation*}
p_{\rm{m}}\> {\rm{tr}}_2\{\Pi\rho'\}-
p_{\rm{n}}\> {\rm{tr}}_2\{\Pi\varrho'\}
\geq p_{\rm{m}}\> {\rm{tr}}_2\{\Pi(\rho'-\varrho')\}
\ .
\end{equation*}
Combining this with (\ref{eq26}) finally gives (\ref{eq22}).
$\blacksquare$

Thus, when the probability $p_{\rm{m}}$ is close to 1, the value of
$D(\rho',\varrho')$ is limited above by a quantity that is
approximately equal to $D(\rho,\varrho)$. In this sense the
considered operations may be related with trace-preserving quantum
operations. For other values of $p_{\rm{m}}$ the upper
bound given by (\ref{eq22}) can appreciably exceed
$D(\rho,\varrho)$. Nevertheless, this bound is nontrivial almost
everywhere. Indeed, under the precondition of Theorem 3 we have
$p_{\rm{m}}-p_{\rm{n}}=D(\rho,\varrho)$. So the right-hand side of
(\ref{eq22}) can be rewritten as $(1-p_{\rm{n}}/p_{\rm{m}})$. If
we represent $p_{\rm{m}}$ along the abscissa and $p_{\rm{n}}$ along
the ordinate then the acceptable values of $p_{\rm{m}}$ and
$p_{\rm{n}}$ lie in the rectangular triangle
$0\leq p_{\rm{n}}<p_{\rm{m}}\leq1$. Except the side $p_{\rm{n}}=0$
of the triangle, the quantity $(1-p_{\rm{n}}/p_{\rm{m}})$ is less
than 1 and the bound given by (\ref{eq22}) is therefore nontrivial.

To each point $(p_{\rm{m}},p_{\rm{n}})$ of the triangle assign
normalized inputs $\rho$ and $\varrho$ such that
$p_{\rm{m}}-p_{\rm{n}}=D(\rho,\varrho)$,
$p_{\rm{m}}={\rm{tr}}_2\{{\cal{E}}(\rho)\}$ and
$p_{\rm{n}}={\rm{tr}}_2\{{\cal{E}}(\varrho)\}$ for given maximizing
operation ${\cal{E}}$. Desired states are defined by (\ref{eq18}) and
(\ref{eq19}), when both the $\lambda_q$'s and the $\varkappa_r$'s sum
to $(p_{\rm{m}}-p_{\rm{n}})$ and the right-hand side of (\ref{eq20})
is equal to $(1-p_{\rm{m}}+p_{\rm{n}})$. We shall now consider
$D(\rho',\varrho')$ as a random variable with values from the interval
$[0;1]$. To evaluate average properties of a function of density
matrices, it is necessary to define a certain measure in the set of
considered ones \cite{sommer1}. In general, this is a subject of
independent research. Some statistical properties of random density
matrices have been analyzed by Sommers and Zyczkowski \cite{sommer2}.
Problems of mentioned kind entail the specific tasks, such as
computing the volume of set of mixed states with respect to the chosen
measure \cite{sommer3,sommer4}. A discussion of these questions would
take us to far afield.

Instead, we simply assume that all points of the triangle are
equiprobable. Then the weight of those points that lead to
$D(\rho',\varrho')\leq\xi$ is no less than $\xi$. Indeed, this
inequality is provided by condition $p_{\rm{n}}\geq(1-\xi)p_{\rm{m}}$
together with (\ref{eq22}). So the lower estimate $\xi$ is
obtained as the ratio of areas of two triangles (the first triangle
arises by section of the second one $0\leq
p_{\rm{n}}<p_{\rm{m}}\leq1$ by line
$p_{\rm{n}}=(1-\xi)p_{\rm{m}}$ ). In other words, the probability of
event $D(\rho',\varrho')\leq\xi$ must be no less than $\xi$. The
density equal to 1 corresponds to the probability distribution
equal to $\xi$. By Lemma 2 of Appendix A, the $n$'th--order moment of
$D(\rho',\varrho')$ is no greater than $1/(n+1)$. In particular, the
mean value does not exceed one half. We see that if the quantum
operation maximizes the probability difference between inputs then
the trace distance between outputs must take small values with
significant frequency.

Unlike the trace-preserving quantum operations, the considered
operations may increase the trace distance between two states. But if
the probability difference between these states is maximized by given
operation then a possible growth of the trace distance is limited
above. In such a case the relative increase of the trace distance will
be negligible by several times. Due to (\ref{eq22}), a relative
variation of the trace distance obeys
\begin{equation}
\frac{D(\rho',\varrho')-D(\rho,\varrho)}{D(\rho',\varrho')}
\leq 1-p_{\rm{m}} \ .
\label{eq27}
\end{equation}
We prove (\ref{eq27}) for those pairs of states that satisfy the
equality $d_{\cal{E}}=D$ for the given quantum operation ${\cal{E}}$.
To any such pair we assign a point $(p_{\rm{m}},p_{\rm{n}})$ of the
triangle $0\leq p_{\rm{n}}<p_{\rm{m}}\leq1$. Suppose those points in
which the trace distance increases are uniformly distributed in the
triangle. Estimate the weight of points such that the relative
increase of trace distance is no greater than $\zeta$. This lower
estimate is obtained as the ratio of the trapezoidal area severed by
line $p_{\rm{m}}=1-\zeta$ from the triangle $0\leq
p_{\rm{n}}<p_{\rm{m}}\leq1$ to the whole triangle area. We consider
the relative increase of trace distance as a random variable with
values from the interval $[0;1]$. By calculations, the probability of
the event that the random variable does not exceed $\zeta$ is no less
than $(2\zeta-\zeta^2)$. The latter distribution is assigned to the
density equal to $(2-2\zeta)$. Due to Lemma 2, the $n$'th--order
moment of the random variable does not exceed $2/(n^2+3n+2)$. In
particular, the mean value is less than or equal to one third. Thus,
on the average the relative increase of trace distance is not great.
Such a property seems to be similar to the contractivity under
trace-preserving quantum operations.

There is another characterization of behaviour of the trace
distance under quantum operations maximizing probability difference
between their inputs. In some instances, the formulation in terms
of subnormalized outputs may be more embossed than (\ref{eq22}).
Except the trace-preserving operations, the output ${\cal{E}}(\rho)$
is subnormalized, i.e. ${\rm{tr}}_2\{{\cal{E}}(\rho)\}\leq1$. So an
extension of the notion of trace distance to subnormalized states is
needed. A study of the general case is beyond the scope of this paper.
However, we can give a transparent outline of the case of Hermitian
operators. All the necessary details are gathered in Appendix B. It
is proved there that the trace distance is a metric on the space of
Hermitian operators. We can now establish the desired
characterization.

{\bf Theorem 4}. {\it If the quantum operation ${\cal{E}}$ maximizes
probability difference between normalized states $\rho$ and $\varrho$
then}
\begin{equation}
D({\cal{E}}(\rho),{\cal{E}}(\varrho))
\leq\frac{1}{2}\,D(\rho,\varrho) \ .
\label{eq28}
\end{equation}

{\bf Proof}. We again suppose that
${\rm{tr}}_2\{{\cal{E}}(\rho)\}>{\rm{tr}}_2\{{\cal{E}}(\varrho)\}$.
Due to the precondition of Theorem 4, the difference between these
traces is equal to $D(\rho,\varrho)$. Using this fact and
(\ref{eqb3}), we see that there exists a projector $\Pi$ such that
\begin{equation*}
{\rm{tr}}_2\{\Pi({\cal{E}}(\rho)-{\cal{E}}(\varrho))\}
=D({\cal{E}}(\rho),{\cal{E}}(\varrho))
+\frac{1}{2}\,D(\rho,\varrho) \ .
\end{equation*}
Combining this with (\ref{eq23}), after cancellation we obtain
(\ref{eq28}). $\blacksquare$

Like (\ref{eq22}), in Theorem 4 the nontrivial upper bound on the
trace distance between outputs is established. Namely, if the quantum
operation maximizes probability difference between inputs then the
trace distance between outputs is at most one-half of the trace
distance between inputs. Assume that all the points of the triangle
$0\leq p_{\rm{n}}<p_{\rm{m}}\leq1$ are equiprobable. Then the mean
value of $D(\rho,\varrho)$ is equal to one third. This result is
obtained as the ratio of the integral of $(p_{\rm{m}}-p_{\rm{n}})$
over triangle to the area of triangle. By (\ref{eq28}), the mean
value of $D({\cal{E}}(\rho),{\cal{E}}(\varrho))$ is no greater than
one sixth. Thus, on the average the outputs must be enough close.

We see from (\ref{eq28}) that for examined operations the trace
distance between subnormalized outputs is bounded above when the
inputs form a pair from specified class. At the same time, there is
an important example of opposite behaviour of the trace distance. Let
us consider a procedure of approximate (or probabilistic) duplicating
quantum states called 'quantum cloning' and useful in many tasks of
quantum information processing. Concrete limitations of this
procedure follow from its specification \cite{acin}. After
inspiring paper by Bu\v{z}ek and Hillery \cite{buzek}, the much
various scenarios have been studied --- the deterministic cloning
\cite{bruss,werner,keyl,macchi1} and the probabilistic cloning
\cite{duan1,duan2}, the hybrid scheme \cite{barn2,fiurasek}, the
cloning with prior information \cite{rast5,rast4,qiu2}, applications
to joint measurement of noncommuting observables \cite{sacchi,barn3},
and the tasks \cite{niu,macchi2,cerf} connected with the quantum
cryptography.

Exact clones may be generated by probabilistic process only. Optimal
exact cloning of state secretly chosen from a certain pair of
different pure states $\omega_1$ and $\omega_2$ has the success
probability $1/(1+\Omega)$, where $\Omega$ denotes the fidelity of
(normalized) states $\omega_1$ and $\omega_2$ \cite{duan1,duan2}.
Recall that the fidelity of normalized states $\rho$ and $\varrho$ is
defined by \cite{nchuang,uhlmann1}
\begin{equation}
F(\rho,\varrho):={\rm{tr}}_1\sqrt{\sqrt{\rho}\,\varrho\,\sqrt{\rho}}
\ . \label{eq29}
\end{equation}
[Such a usage of the word 'fidelity' is not unique. In \cite{jozsa}
Jozsa introduced this word for Uhlmann's transition probability
\cite{uhlmann2} equal to square of the right-hand side of
(\ref{eq29}).] For normalized states the trace distance and the
fidelity are related by the inequality
\begin{equation}
D(\rho,\varrho)\leq\sqrt{1-F^2(\rho,\varrho)}
\ , \label{eq30}
\end{equation}
which is always saturated for pure states \cite{nchuang}. The actual
outputs of exact cloning operation ${\cal{G}}$ is expressed as
\begin{equation}
{\cal{G}}(\omega_j)=(1+\Omega)^{-1}\>\omega_j\otimes\omega_j
\ , \label{eq31}
\end{equation}
where $j=1,2$. By multiplicativity of the fidelity \cite{jozsa}, we
have
$F(\omega_1\otimes\omega_1,\omega_2\otimes\omega_2)=F^2(\omega_1,\omega_2)
=\Omega^2$. Since both the states $\omega_j$ and
$\omega_j\otimes\omega_j$ are pure, the equality in (\ref{eq30})
holds whence
\begin{equation*}
D(\omega_1\otimes\omega_1,\omega_2\otimes\omega_2)=
\sqrt{1+\Omega^2}\>D(\omega_1,\omega_2) \ .
\end{equation*}
Using the customary definition of the trace distance, equation
(\ref{eq31}) and the last relation, we then obtain
\begin{equation}
D({\cal{G}}(\omega_1),{\cal{G}}(\omega_2))=
\frac{\sqrt{1+\Omega^2}}{1+\Omega}\>D(\omega_1,\omega_2)
\ . \label{eq32}
\end{equation}
Because the normalized states $\omega_1$ and $\omega_2$ are
different, a value of $\Omega=F(\omega_1,\omega_2)$ lies in the
interval $[0;1)$. For such values the multiplier of
$D(\omega_1,\omega_2)$ in (\ref{eq32}) is decreasing function of
$\Omega$ and, therefore, is greater than $1/\sqrt{2}$. Thus, if the
quantum operation ${\cal{G}}$ is designed to clone exactly the
prescribed pure states $\omega_1$ and $\omega_2$ then
\begin{equation}
D({\cal{G}}(\omega_1),{\cal{G}}(\omega_2))>
\frac{1}{\sqrt{2}}\>D(\omega_1,\omega_2)
\ . \label{eq33}
\end{equation}

Let us compare the two results established by equations (\ref{eq28})
and (\ref{eq33}) respectively. The similarity is that each of these
results imposes some bound on the trace distance between two outputs
when the two input states form specified pair. The differences
are significant in the following respects. First, the trace distance
between outputs of considered operation ${\cal{E}}$ is bounded above,
the trace distance between outputs of exact cloning operation
${\cal{G}}$ is bounded below. Second, inequality (\ref{eq28}) is
valid for infinitely many pairs of inputs, inequality (\ref{eq33}) is
valid for only one pair of inputs. The more demonstrative of the two
differences is the first. The second difference is rather a
manifestation of the fact that in physical processes a loss of
distinguishability usually occurs.

\protect\section{Relations with sine distance}

In this section, we shall discuss a relationship of the trace distance
and a close measure that is called 'sine distance' in \cite{rast1}.
There are the two useful definitions of the sine distance. The first
definition is based on the concept of purifications and the notion of
angle between quantum states. In \cite{rast2} the angle
$\Delta(\rho,\varrho)\in[0;\pi/2]$ between states $\rho$ and
$\varrho$ has been defined by
\begin{equation*}
\Delta(\rho,\varrho):=\underset{|\Phi\rangle,|\Psi\rangle}{\min}
\Delta(|\Phi\rangle,|\Psi\rangle) \ ,
\end{equation*}
where the minimization is over all purifications $|\Phi\rangle$ of
$\rho$ and $|\Psi\rangle$ of $\varrho$, and
$\Delta(|\Phi\rangle,|\Psi\rangle):=\arccos|\langle\Phi|\Psi\rangle|$.
The sine distance between states $\rho$ and $\varrho$ is then defined
as \cite{rast1}
\begin{equation}
C(\rho,\varrho):=\sin\Delta(\rho,\varrho)
\ . \label{eq34}
\end{equation}
The name 'sine distance' has been arisen from (\ref{eq34}).
According to the second definition \cite{rast1}, the sine distance
$C(\rho,\varrho)$ is defined as the right-hand side of
(\ref{eq30}). These definitions are consistent, because there holds
\cite{nchuang}
\begin{equation}
F(\rho,\varrho)=\cos\Delta(\rho,\varrho)
\ . \label{eq35}
\end{equation}
It turned out that the sine distance is useful in the state-dependent
quantum cloning. Following \cite{bruss}, state-dependent cloners are
usually evaluated with respect to those figures of merit that are
based on the fidelity. In \cite{rast4,rast3} the new figure of
merit, based on the sine distance and called 'relative error', has
been proposed. A study of cloners with respect to the relative error
has allowed us to complete the portrait of state-dependent cloning
\cite{rast3}. In addition, the considered distance seems to be
useful in the context of quantum computation \cite{nielsen}.

If both the states are pure, the equality in (\ref{eq30}) takes
place and, therefore, the sine distance is equal to the trace
distance. In general, however, the sine distance can be larger than
the trace distance. Consider the pure state $|0\rangle\langle 0|$ and
the mixed state $\varrho$ with spectral decomposition
\begin{equation*}
\varrho=(1-\lambda)\,|0\rangle\langle 0|
+\sum\nolimits_{r\not=0} \varkappa_r
\> |r\rangle\langle r| \ .
\end{equation*}
It is easy to check that
$F(|0\rangle,\varrho)=\sqrt{1-\lambda}$, whence
$C(|0\rangle,\varrho)=\sqrt{\lambda}$. Splitting operator
$(|0\rangle\langle 0|-\varrho)$ into positive and negative parts is
obvious, and from (\ref{eq11}) we obtain
$D(|0\rangle,\varrho)=\lambda$. The maximum of function
$\sqrt{\lambda}-\lambda=1/4-(\sqrt{\lambda}-1/2)^2$ is equal to one
forth and reached at $\lambda=1/4$. So for this value of $\lambda$ we
have
\begin{equation}
C(|0\rangle,\varrho)-D(|0\rangle,\varrho)=1/4
\ . \label{eq36}
\end{equation}
%We shall now give the lower bound and the upper bound on the maximum
%of difference between the sine distance and the trace distance.

{\bf Theorem 5.} {\it The maximum of difference between the sine
distance and the trace distance satisfies}
\begin{equation}
\frac{1}{4}\leq \underset{\rho,\varrho}{\max}
\,\{C(\rho,\varrho)-D(\rho,\varrho)\}\leq\sqrt{2}-1
\ , \label{eq37}
\end{equation}
{\it where the maximization is over all states $\rho$ and $\varrho$.}

{\bf Proof.} The lower bound follows from (\ref{eq36}). As is
shown in \cite{nchuang,fuchs},
$1-F(\rho,\varrho)\leq D(\rho,\varrho)$ whence
\begin{equation*}
C(\rho,\varrho)-D(\rho,\varrho)\leq
C(\rho,\varrho)+F(\rho,\varrho)-1 \ .
\end{equation*}
Due to (\ref{eq34}) and (\ref{eq35}), the last inequality can be
rewritten as
\begin{equation}
C(\rho,\varrho)-D(\rho,\varrho)\leq
\sin\Delta(\rho,\varrho)+\cos\Delta(\rho,\varrho)-1
\ . \label{eq38}
\end{equation}
The upper bound is provided by (\ref{eq38}) and Lemma 3 of
Appendix A. $\blacksquare$

It is not insignificant that in the case of single qubits the lower
bound in (\ref{eq37}) is saturated. In other words, the maximum
of difference between the sine distance and the trace distance is
equal to one forth. As always, we represent the density matrices by
$\rho=(1/2)\left\{{\mathbf{1}}+{\vec{u}}\cdot{\vec{\sigma}}\right\}$
and
$\varrho=(1/2)\left\{{\mathbf{1}}+{\vec{v}}\cdot{\vec{\sigma}}\right\}$.
Here ${\vec{u}}$ and ${\vec{v}}$ are Bloch vectors and
${\vec{\sigma}}$ denotes the three-component vector of Pauli
matrices. The square of the fidelity of states $\rho$ and $\varrho$ is
then expressed as \cite{jozsa}
\begin{equation*}
F^2(\rho,\varrho)=\frac{1}{2}
\left\{1+{\vec{u}}\cdot{\vec{v}}
+\sqrt{1-u^2}\,\sqrt{1-v^2}\right\} \ .
\end{equation*}
Next, the trace distance between two single qubit states is equal to
one-half of modulus of difference between their Bloch vectors
\cite{nchuang}. So the difference between the sine distance and the
trace distance is equal to the function
\begin{align*}
f(u,v,\eta) =&\,\frac{1}{\sqrt{2}}\left\{1-uv\eta
-\sqrt{1-u^2}\,\sqrt{1-v^2}\right\}^{1/2} \\
&-\frac{1}{2} \left\{u^2+v^2-2uv\eta\right\}^{1/2} \ ,
\end{align*}
where $\eta$ denotes the cosine of angle between ${\vec{u}}$ and
${\vec{v}}$. Acceptable values of variables $u$, $v$ and $\eta$ lie
in the parallelepiped defined by $0\leq u\leq1$, $0\leq v\leq1$ and
$-1\leq\eta\leq1$. Finding maximum of the function $f(u,v,\eta)$ in
the parallelepiped is a task of elementary calculus. It has been
verified that desired maximum is equal to one forth. But we refrain
from presenting the calculations here.

To sum up we see that the trace distance is closely related to the
sine distance. Moreover, in the case of pure states the two distance
measures are equal to each other. In general, the sine distance can
be larger than the trace distance. So the trace distance is sometimes
tighter. But the maximum of difference between the sine distance and
the trace distance lies between values $1/4$ and $(\sqrt{2}-1)$. The
former takes place in the case of single qubits. It would be
interesting to study a dependence of this maximum on the
dimensionality of state space. But this problem seems to be enough
difficult.

\protect\section{Conclusion}

We have considered the trace distance from the viewpoint of quantum
operation formalism. The new definition of trace distance in terms of
a maximum over all quantum operations was proposed. The definition
proposed in this paper has the advantage of a physical interpretation
of the trace distance in terms of quantum operations. In connection
with this definition the interesting subclass of maximizing quantum
operation was specified. It has been shown that each of such
operations maximizes a difference between two probabilities generated
by the operation on some pairs of inputs. For each pair of different
states there exist an uncountably infinite number of quantum
operations with specified property. Conversely, for each quantum
operation of described type there exist an uncountably infinite
number of pairs of those states that probability difference between
them is maximized by the operation.

It turned out that if quantum operation maximizes the probability
difference between inputs then the trace distance between outputs
is bounded above. Due to made estimates of trace distance between
outputs, described operations have been related to the
trace-preserving quantum operations. The revealed property seems
to be similar to the well-known contractivity under the
trace-preserving quantum operations. But this property is valid
only for specific pairs of inputs. Finally, we have discussed
relations of the trace distance to a measure called 'sine
distance'. The lower and upper bounds on the maximum of difference
between the sine distance and the trace distance were obtained. In
the case of single qubits the exact value of this maximum is
mentioned. These results show that the sine distance and the trace
distance are closely related.

\newpage
\appendix

\protect\section{Three lemmas}

Let us consider a product of two positive operators, one of which is
fixed and other of which is freely variable. We find the maximal and
minimal values of the trace of this product. By $\Theta$ and
$\theta$, we denote the maximal and minimal eigenvalues of the fixed
positive operator ${\mathbf{T}}$, respectively. Then the following
statement takes place.

{\bf Lemma 1.} {\it For the given positive operator ${\mathbf{T}}$,}
\begin{align}
\underset{{\rm{tr}}({\mathbf{Q}})={\mathfrak{D}}}{\max}
{\rm{tr}}({\mathbf{TQ}})&=\Theta\cdot{\mathfrak{D}} \ ,
\label{eqa01} \\
\underset{{\rm{tr}}({\mathbf{Q}})={\mathfrak{D}}}{\min}
{\rm{tr}}({\mathbf{TQ}})&=\theta\cdot{\mathfrak{D}} \ ,
\label{eqa02}
\end{align}
{\it where both the maximization and minimization is over all
positive operators ${\mathbf{Q}}$ satisfying
${\rm{tr}}({\mathbf{Q}})={\mathfrak{D}}$.}

{\bf Proof.} Using the spectral decomposition of operator
${\mathbf{T}}$ and the definitions of $\theta$ and $\Theta$, we obtain
that for each normalized state $|q\rangle$
\begin{equation}
\theta\leq\langle q|{\mathbf{T}}|q\rangle\leq\Theta
\ . \label{eqa03}
\end{equation}
Due to the properties of the trace and (\ref{eq6}),
${\rm{tr}}({\mathbf{TQ}})=\sum\nolimits_{q}\lambda_q\>\langle
q|{\mathbf{T}}|q\rangle$. This, when combined with (\ref{eqa03}),
finally gives
\begin{equation}
\theta\cdot{\mathfrak{D}}\leq {\rm{tr}}({\mathbf{TQ}})
\leq\Theta\cdot{\mathfrak{D}} \ . \label{eqa04}
\end{equation}
Here we used that
${\rm{tr}}({\mathbf{Q}})=\sum_{q}\lambda_q={\mathfrak{D}}$. To reach
the lower bound in (\ref{eqa04}) we take a nontrivial subspace of the
eigenspace of ${\mathbf{T}}$ corresponding to eigenvalue $\theta$;
then ${\mathbf{Q}}$ should be the projector onto this subspace
multiplied by the ratio of ${\mathfrak{D}}$ to trace of the
projector. To reach the upper bound in (\ref{eqa04}) we take a
nontrivial subspace of the eigenspace of ${\mathbf{T}}$ corresponding
to eigenvalue $\Theta$; then ${\mathbf{Q}}$ should be the projector
onto that subspace multiplied by the ratio of ${\mathfrak{D}}$ to
trace of the projector. $\blacksquare$

Let $X$ and $Y$ be the real-valued random variables with probability
densities $g(x)$ and $h(y)$ respectively. It is sufficient for our
aims to consider only those probability densities that vanish outside
a certain interval $[0;R]$. A distribution function of $\xi$ is
defined as the probability that a value of the random variable is no
greater than $\xi$ \cite{fell}. This function is obtained by
integration from $0$ to $\xi$ of corresponding probability density.
The moments are important quantitative indices of distribution
properties \cite{fell}. In our case the $n$'th--order moments of $X$
and $Y$ are expressed by
\begin{align}
\langle X^n\rangle&=\int_{0}^{R}
 x^n\,g(x) \,dx \ ,
\label{eqa1} \\
\langle Y^n\rangle&=\int_{0}^{R}
 y^n\,h(y) \,dy \ .
\label{eqa2}
\end{align}
We shall now show that if the two distribution functions satisfy the
same inequality for all $\xi$ in $[0;R]$, then the two moments of
$n$'th order satisfy the opposite inequality.

{\bf Lemma 2.} {\it If there holds
$\int_{0}^{\xi}g(x)\,dx\geq\int_{0}^{\xi}h(y)\,dy$ for all
$\xi\in[0;R]$ then}
\begin{equation}
\langle X^n\rangle\leq\langle Y^n\rangle \qquad (n>0)\ .
\label{eqa3}
\end{equation}

{\bf Proof.} The quantity $ny^{n-1}\int_{0}^{y}\{g(x)-h(x)\}\,dx$ is
nonnegative for all $y\in[0;R]$ due to the precondition of Lemma 2.
So by integration from $y=0$ to $y=R$ of this nonnegative quantity we
obtain
\begin{align}
 & \int_{0}^{R} dy \int_{0}^{y} dx\,ny^{n-1}\{g(x)-h(x)\}
\nonumber\\
 & =\int_{0}^{R} dx \int_{x}^{R} dy\,ny^{n-1}\{g(x)-h(x)\}
\nonumber\\
 & =\int_{0}^{R} dx\,(R^n-x^n)\{g(x)-h(x)\}\geq 0
\ . \label{eqa4}
\end{align}
In the last line of (\ref{eqa4}) the multiplier of $R^n$ is zero
by the normalization of densities. Combining this with (\ref{eqa1})
and (\ref{eqa2}) finally gives (\ref{eqa3}).  $\blacksquare$

It should be noted that the above result remains valid when the
probability densities are distributed among the whole positive
semiaxis. To prove this we must consider the limit $R\to+\infty$.
It turns out that if the integrals in (\ref{eqa1}) and (\ref{eqa2})
are convergent then the statement of Lemma 2 is still correct. We do
not enter into details here because in Section 3 we deal with
probability densities concentrated on the interval $[0;1]$. In
general, Lemma 2 can be extended to any function of the random
variable such that its derivative is nonnegative in those intervals
on which the densities are concentrated. A discussion of this
question would be out of the place here.

{\bf Lemma 3.} {\it For arbitrary angle $\alpha$ there holds}
\begin{equation}
\sin\alpha+\cos\alpha\leq\sqrt{2}
\ . \label{eqa5}
\end{equation}

{\bf Proof.} By doing usual trigonometry, we obtain
\begin{align}
\sin\alpha+\cos\alpha
&= \sqrt{2}\>\bigl(\sin\alpha\cos(\pi/4)+
\cos\alpha\sin(\pi/4) \bigr) \nonumber \\
&= \sqrt{2}\>\sin(\alpha+\pi/4) \ . \nonumber
\end{align}
Because the sine does not exceed one, this equality provides
(\ref{eqa5}). $\blacksquare$

\protect\section{Trace distance between \\ Hermitian operators}

In general, the right-hand side of (\ref{eq10}) can naturally be
extended in much broad context. Indeed, the expression for trace
distance between two density operators is regardless of the
normalization and the positivity of them. We shall restrict our
consideration to the case of Hermitian operators. In the first place,
this subclass of operators is extremely important. In the second
place, under such a restriction we can give a simple analysis of the
properties of the trace distance. The trace distance between Hermitian
operators ${\mathsf{A}}$ and ${\mathsf{B}}$ is defined by
\begin{equation}
D({\mathsf{A}},{\mathsf{B}}):=\frac{1}{2}\ {\rm{tr}}\,
|{\mathsf{A}}-{\mathsf{B}}| \ . \label{eqb1}
\end{equation}
Due to Hermiticity of ${\mathsf{A}}$ and ${\mathsf{B}}$ we can obtain
a direct analogue of (\ref{eq11}). Like a difference between
density matrices, Hermitian operator $({\mathsf{A}}-{\mathsf{B}})$
can be written as
${\mathsf{A}}-{\mathsf{B}}={\mathbf{P}}-{\mathbf{S}}$, where
${\mathbf{P}}$ and ${\mathbf{S}}$ are positive operators with
orthogonal supports. These operators are got from the spectral
decomposition of $({\mathsf{A}}-{\mathsf{B}})$ by the same way that
leads to (\ref{eq6}) and (\ref{eq7}). Drawing analogy with
(\ref{eq11}), we immediately obtain
\begin{equation}
D({\mathsf{A}},{\mathsf{B}})=\frac{1}{2}\ {\rm{tr}}({\mathbf{P}})
+\frac{1}{2}\ {\rm{tr}}({\mathbf{S}})\ . \label{eqb2}
\end{equation}
In contrast to the case of normalized density operators, neither
${\rm{tr}}({\mathbf{P}})$ nor ${\rm{tr}}({\mathbf{S}})$ are equal to
the right-hand side of (\ref{eqb2}) (except when
${\rm{tr}}({\mathsf{A}})={\rm{tr}}({\mathsf{B}})$ solely).

The distance defined by (\ref{eqb1}) is just a metric on the
space of Hermitian operators. It is obvious that the distance takes
nonnegative real values, that $D({\mathsf{A}},{\mathsf{B}})=0$ if and
only if ${\mathsf{A}}={\mathsf{B}}$, and that
$D({\mathsf{A}},{\mathsf{B}})=D({\mathsf{B}},{\mathsf{A}})$. The only
vague step is a proof of the triangle inequality. Here a
generalization of (\ref{eq26}) is needed.

{\bf Lemma 4.} {\it For arbitrary two Hermitian operators
${\mathsf{A}}$ and ${\mathsf{B}}$}
\begin{equation}
\underset{\Pi\leq{\mathbf{1}}}{\max}
\ {\rm{tr}}\{\Pi({\mathsf{A}}-{\mathsf{B}})\}=
D({\mathsf{A}},{\mathsf{B}})+\frac{{\rm{tr}}({\mathsf{A}})-
{\rm{tr}}({\mathsf{B}})}{2} \ , \label{eqb3}
\end{equation}
{\it where maximum is taken over all positive operators $\Pi$
satisfying $\Pi\leq{\mathbf{1}}$ (or alternately over all
projectors).}

{\bf Proof.} Taking the trace of operator
${\mathsf{A}}-{\mathsf{B}}={\mathbf{P}}-{\mathbf{S}}$ and using
(\ref{eqb2}), we obtain
\begin{equation*}
D({\mathsf{A}},{\mathsf{B}})+\frac{1}{2}\,[{\rm{tr}}
({\mathsf{A}})-{\rm{tr}}({\mathsf{B}})]
={\rm{tr}}({\mathbf{P}})\ .
\end{equation*}
Prove that the left-hand side of (\ref{eqb3}) is
equal to ${\rm{tr}}({\mathbf{P}})$. For any positive operator
$\Pi\leq{\mathbf{1}}$ there holds
\begin{equation*}
{\rm{tr}}\{\Pi({\mathsf{A}}-{\mathsf{B}})\}=
{\rm{tr}}\{\Pi({\mathbf{P}}-{\mathbf{S}})\}
\leq{\rm{tr}}\{\Pi{\mathbf{P}}\}
\leq{\rm{tr}}({\mathbf{P}}) \ .
\end{equation*}
When $\Pi$ is the projector onto the support of ${\mathbf{P}}$, both
the last inequalities are saturated. $\blacksquare$

Note that Lemma 4 is related in kinship to Theorem 1. In
(\ref{eqb3}) the maximization is over all positive operators $\Pi$
meeting $\Pi\leq{\mathbf{1}}$. If we substitute the defined by
(\ref{eq4}) operator ${\mathbf{T}}$ for abstract $\Pi$ then in the
left-hand side of (\ref{eqb3}) we obtain the maximum over all quantum
operations. In this sense, the statement of Theorem 1 provides a kind
of physical interpretation of (\ref{eqb3}) for the case of density
operators. Besides, in Theorem 1 the explicit conditions of
achievement of the maximum are established. On other hand, Lemma 4
deals with arbitrary Hermitian operators. Furthermore, its
applications to the proof of the triangle inequality and the
convexity do not involve conditions of maximum achievement. We now
note from (\ref{eqb3}) that there exists a projector $\Pi$ such that
\begin{equation}
{\rm{tr}}\{\Pi({\mathsf{A}}-{\mathsf{B}})\}-\frac{1}{2}\,
[{\rm{tr}}({\mathsf{A}})-{\rm{tr}}({\mathsf{B}})]
=D({\mathsf{A}},{\mathsf{B}})
 \ . \label{eqb4}
\end{equation}
In accordance with Lemma 4, we further have
\begin{align}
{\rm{tr}}\{\Pi({\mathsf{A}}-{\mathsf{C}})\}-\frac{1}{2}\,
[{\rm{tr}} ({\mathsf{A}})-{\rm{tr}}({\mathsf{C}})] & \leq
D({\mathsf{A}},{\mathsf{C}}) \ , \nonumber\\
{\rm{tr}}\{\Pi({\mathsf{C}}-{\mathsf{B}})\}-\frac{1}{2}\,
[{\rm{tr}} ({\mathsf{C}})-{\rm{tr}}({\mathsf{B}})] & \leq
D({\mathsf{C}},{\mathsf{B}}) \ . \nonumber
\end{align}
Summing the two last inequalities and using (\ref{eqb4}), we
finally obtain that $D({\mathsf{A}},{\mathsf{B}})\leq
D({\mathsf{A}},{\mathsf{C}})+D({\mathsf{C}},{\mathsf{B}})$. Thus, the
triangle inequality holds too.

The trace distance between density matrices satisfies the following
two properties: the joint convexity and the convexity \cite{nchuang}.
These properties remain valid for Hermitian matrices. Let $\{p_j\}$
be probability distribution, and ${\mathsf{A}}_j$ and
${\mathsf{B}}_j$ be Hermitian operators with labels from the same
set. Then
\begin{equation}
D\left(\sum\nolimits_j p_j {\mathsf{A}}_j\>,\sum\nolimits_j p_j
{\mathsf{B}}_j\right) \leq \sum\nolimits_j p_j
D({\mathsf{A}}_j,{\mathsf{B}}_j) \ ,
\label{eqb5}
\end{equation}
that is the trace distance is jointly convex in its inputs.
Substituting ${\mathsf{C}}$ for all ${\mathsf{B}}_j$'s into
(\ref{eqb5}) and using the condition $\sum_j p_j=1$, we obtain
\begin{equation*}
D\left(\sum\nolimits_j p_j {\mathsf{A}}_j\>,{\mathsf{C}}\right)
\leq \sum\nolimits_j p_j D({\mathsf{A}}_j,{\mathsf{C}}) \ .
\end{equation*}
That is, the trace distance is convex function on the set of
Hermitian matrices.

The proof of (\ref{eqb5}) is simple. By ${\mathsf{A}}$ and
${\mathsf{B}}$ we denote $\sum_jp_j{\mathsf{A}}_j$ and
$\sum_jp_j{\mathsf{B}}_j$ respectively. Due to (\ref{eqb3}) there
exists a projector $\Pi$ such that
\begin{align}
D({\mathsf{A}},{\mathsf{B}}) &=
{\rm{tr}}\{\Pi({\mathsf{A}}-{\mathsf{B}})\}
-\frac{1}{2}\,[{\rm{tr}}({\mathsf{A}})-{\rm{tr}}({\mathsf{B}})]
\nonumber\\
&=\sum\nolimits_j p_j
\,{\rm{tr}}\{\Pi({\mathsf{A}}_j-{\mathsf{B}}_j)\}
-\frac{1}{2}\,[{\rm{tr}}({\mathsf{A}})-{\rm{tr}}({\mathsf{B}})]
\nonumber\\
&\leq\sum\nolimits_j p_j \left\{D({\mathsf{A}}_j,{\mathsf{B}}_j)
+\frac{1}{2}\,[{\rm{tr}}({\mathsf{A}}_j)-{\rm{tr}}({\mathsf{B}}_j)]
\right\}
\nonumber\\
&\
-\frac{1}{2}\,[{\rm{tr}}({\mathsf{A}})-{\rm{tr}}({\mathsf{B}})] \
. \label{eqb6}
\end{align}
Here in the last part of (\ref{eqb6}) the statement of Lemma 4
was applied. After cancellation in this part we obtain
(\ref{eqb5}).


\begin{thebibliography}{99}

\bibitem{kraus1}%-----------------------------------------------------
Kraus K 1983 {\it States, Effects and Operations: Fundamental Notions
of Quantum Theory} ({\it Lecture Notes in Physics} vol. 190) (Berlin:
Springer-Verlag)

\bibitem{nchuang}%-----------------------------------------------
Nielsen M A and Chuang I L 2000 {\it Quantum Computation and Quantum
Information} (Cambridge: Cambridge University Press)

\bibitem{hell1}%-----------------------------------------------------
Hellwig K-E and Kraus K 1969 {\it Commun. Math. Phys.} {\bf 11} 214

\bibitem{hell2}%-----------------------------------------------------
Hellwig K-E and Kraus K 1970 {\it Commun. Math. Phys.} {\bf 16} 142

\bibitem{kraus2}%-----------------------------------------------------
Kraus K 1971 {\it Ann. Phys.} {\bf 64} 311

\bibitem{lind}%-----------------------------------------------------
Lindblad G 1975 {\it Commun. Math. Phys.} {\bf 40} 147

\bibitem{choi}%-----------------------------------------------------
Choi M-D 1975 {\it Linear Algebra Appl.} {\bf 10} 285

\bibitem{bennett}%--------------------------------------------------
Bennett C H 1992 {\it Phys. Rev. Lett.} {\bf 68} 3121

\bibitem{paris}%------------------------------------------------
Olivares S and Paris M G A 2004 {\it J. Opt. B: Quantum Semiclass.
Opt.} {\bf 6} 69

\bibitem{helstrom}%-----------------------------------------------
Helstrom C W 1976 {\it Quantum Detection and Estimation Theory}
(New York: Academic Press)

\bibitem{ivan}%-----------------------------------------------
Ivanovic I D 1987 {\it Phys. Lett.} A {\bf 123} 257

\bibitem{dieks}%-----------------------------------------------
Dieks D 1988 {\it Phys. Lett.} A {\bf 126} 303

\bibitem{peres1}%-----------------------------------------------
Peres A 1988 {\it Phys. Lett.} A {\bf 128} 19

\bibitem{chef1}%-----------------------------------------------
Chefles A 1998 {\it Phys. Lett.} A {\bf 239} 339

\bibitem{chef2}%-----------------------------------------------
Chefles A and Barnett S M 1998 {\it Phys. Lett.} A {\bf 250} 223

\bibitem{turner}%-----------------------------------------------
Rudolph T, Spekkens R W and Turner P S 2003 {\it Phys. Rev.} A
{\bf 68} 010301(R)

\bibitem{peres2}%-----------------------------------------------
Peres A and Terno D R 1998 {\it J. Phys. A: Math. Gen.} {\bf 34} 7105

\bibitem{barnum}%-----------------------------------------------
Barnum H, Caves C M, Fuchs C A, Jozsa R and Schumacher B 1996
{\it Phys. Rev. Lett.} {\bf 76} 2818

\bibitem{caves}%------------------------------------------------
Nielsen M A and Caves C M 1997 {\it Phys. Rev.} A {\bf 55} 2547

\bibitem{barn1}%-----------------------------------------------
Chefles A and Barnett S M 1998 {\it J. Phys. A: Math. Gen.} {\bf 31}
10097

\bibitem{qiu1}%-----------------------------------------------
Qiu D 2002 {\it J. Phys. A: Math. Gen.} {\bf 35} 6931

\bibitem{feng}%-----------------------------------------------
Feng Y, Duan R and Ji Z 2005 {\it Phys. Rev.} A {\bf 72} 012313

\bibitem{jezek}%-----------------------------------------------
Fiura\v{s}ek J and Je\v{z}ek M 2003 {\it Phys. Rev.} A {\bf 67} 012321

\bibitem{aharonov}%--------------------------------------------
Aharonov D, Kitaev A and Nisan N 1998 Quantum circuits with mixed
states {\it Preprint} quant-ph/9806029

\bibitem{reed}%-------------------------------------------------
Reed M and Simon B 1972 {\it Methods of Modern Mathematical
Physics} vol. 1 {\it Functional Analysis} (New York: Academic
Press)

\bibitem{rast1}%-----------------------------------------------
Rastegin A E 2006 Sine distance for quantum states {\it Preprint}
quant-ph/0602112

\bibitem{ruskai}%-----------------------------------------------
Ruskai M B 1994 {\it Rev. Math. Phys.} {\bf 6} 1147

\bibitem{sommer1}%-----------------------------------------------
Zyczkowski K and Sommers H-J 2001 {\it J. Phys. A: Math. Gen.}
{\bf 34} 7111

\bibitem{sommer2}%-----------------------------------------------
Sommers H-J and Zyczkowski K 2004 {\it J. Phys. A: Math. Gen.}
{\bf 37} 8457

\bibitem{sommer3}%-----------------------------------------------
Sommers H-J and Zyczkowski K 2003 {\it J. Phys. A: Math. Gen.}
{\bf 36} 10083

\bibitem{sommer4}%-----------------------------------------------
Zyczkowski K and Sommers H-J 2003 {\it J. Phys. A: Math. Gen.}
{\bf 36} 10115

\bibitem{acin}%-----------------------------------------------
Scarani V, Iblisdir S, Gisin N and Acin A 2005 {\it Rev. Mod. Phys.}
{\bf 77} 1225

\bibitem{buzek}%-----------------------------------------------
Bu\v{z}ek V and Hillery M 1996 {\it Phys. Rev.} A {\bf 54} 1844

\bibitem{bruss}%------------------------------------------------
Bru{\ss} D, DiVincenzo D P, Ekert A, Fuchs C A, Macchiavello C
and Smolin J A 1998 {\it Phys. Rev.} A {\bf 57} 2368

\bibitem{werner}%---------------------------------------------
Werner R F 1998 {\it Phys. Rev.} A {\bf 58} 1827

\bibitem{keyl}%-----------------------------------------------
Keyl M and Werner R F 1999 {\it J. Math. Phys.} {\bf 40} 3283

\bibitem{macchi1}%------------------------------------------------
Macchiavello C 2000 {\it J. Opt. B: Quantum Semiclass. Opt.} {\bf 2}
144

\bibitem{duan1}%-----------------------------------------------
Duan L-M and Guo G-C 1998 {\it Phys. Rev. Lett.} {\bf 80} 4999

\bibitem{duan2}%-----------------------------------------------
Duan L-M and Guo G-C 1998 {\it Phys. Lett.} A {\bf 243} 261

\bibitem{barn2}%---------------------------------------------
Chefles A and Barnett S M 1999 {\it Phys. Rev.} A {\bf 60} 136

\bibitem{fiurasek}%---------------------------------------------
Fiura\v{s}ek J 2004 {\it Phys. Rev.} A {\bf 70} 032308

\bibitem{rast5}%-----------------------------------------------
Rastegin A E 2003 {\it Phys. Rev.} A {\bf 68} 032303

\bibitem{rast4}%-----------------------------------------------
Rastegin A E 2003 {\it J. Opt. B: Quantum Semiclass. Opt.} {\bf 5}
S647

\bibitem{qiu2}%-----------------------------------------------
Qiu D 2006 {\it J. Phys. A: Math. Gen.} {\bf 39} 5135

\bibitem{sacchi}%------------------------------------------------
D'Ariano G M, Macchiavello C and Sacchi M F 2001
{\it J. Opt. B: Quantum Semiclass. Opt.} {\bf 3} 44

\bibitem{barn3}%---------------------------------------------
Brougham T, Andersson E and Barnett S M 2006 {\it Phys. Rev.} A
{\bf 73} 062319

\bibitem{niu}%------------------------------------------------
Niu C-S and Griffiths R B 1999 {\it Phys. Rev.} A {\bf 60} 2764

\bibitem{macchi2}%------------------------------------------------
Bru{\ss} D and Macchiavello C 2001 {\it J. Phys. A: Math. Gen.}
{\bf 34} 6815

\bibitem{cerf}%------------------------------------------------
Cerf N J, Bourennane M, Karlsson A and Gisin N 2002 {\it Phys. Rev.
Lett.} {\bf 88} 127902

\bibitem{uhlmann1}%---------------------------------------------
Uhlmann A 2000 {\it Rep. Math. Phys.} {\bf 45} 407

\bibitem{jozsa}%-----------------------------------------------
Jozsa R 1994 {\it J. Mod. Opt.} {\bf 41} 2315

\bibitem{uhlmann2}%---------------------------------------------
Uhlmann A 1976 {\it Rep. Math. Phys.} {\bf 9} 273

\bibitem{rast2}%-----------------------------------------------
Rastegin A E 2003 {\it Phys. Rev.} A {\bf 67} 012305

\bibitem{rast3}%-----------------------------------------------
Rastegin A E 2002 {\it Phys. Rev.} A {\bf 66} 042304

\bibitem{nielsen}%----------------------------------------------
Gilchrist A, Langford N K and Nielsen M A 2005 {\it Phys. Rev.} A
{\bf 71} 062310

\bibitem{fuchs}%-----------------------------------------------
Fuchs C A and van de Graaf J 1999 {\it IEEE Trans. Inf. Theory}
{\bf 45} 1216

\bibitem{fell}%-----------------------------------------------
Feller W 1971 {\it An Introduction to Probability Theory and its
Applications} vol. II (New York: Wiley)

\end{thebibliography}
\end{document}